# Licensed-Assisted Access to Unlicensed Spectrum in LTE Release 13

Hwan-Joon (Eddy) Kwon, Jeongho Jeon, Abhijeet Bhorkar, and Qiaoyang Ye, Intel Corporation

Hiroki Harada, Yu Jiang, Liu Liu, and Satoshi Nagata, NTT DOCOMO

Boon Loong Ng, Thomas Novlan, Jinyoung Oh, and Wang Yi, Samsung Electronics

*Abstract*—Exploiting the unlicensed spectrum is considered by 3GPP as one promising solution to meet the ever-increasing traffic growth. As a result, one major enhancement for LTE in Release 13 has been to enable its operation in the unlicensed spectrum via Licensed-Assisted Access (LAA). In this article, we provide an overview of the Release 13 LAA technology including motivation, use cases, LTE enhancements for enabling the unlicensed band operation, and the coexistence evaluation results contributed by 3GPP participants.

I. INTRODUCTION

To cope with ever-increasing traffic demand, the 3$^{rd}$ Generation Partnership Project (3GPP) has been continuously endeavoring to increase the network capacity by improving the spectral efficiency of the Long Term Evolution (LTE) system through the introduction of higher order modulations, advanced multi-input multi-output (MIMO) antenna technologies, and multi-cell coordination techniques, to name a few. Another fundamental way to improve the network capacity is to expand the system bandwidth, but newly available spectrum in the lower frequency bands, which have traditionally been individually allocated to each mobile network operator, has become very scarce. This is the main rationale behind the recent study item (SI) and work item (WI) in 3GPP Release 13 to enable the operation of an LTE system in unlicensed spectrum. Since 3GPP considers unlicensed spectrum as supplemental to licensed spectrum, this new feature is called Licensed-Assisted Access (LAA) to unlicensed spectrum, often referred to as LAA. One important consideration for operating LTE in unlicensed spectrum is to ensure fair coexistence with the incumbent systems such as the Wireless Local Area Networks (WLANs)[1], which has been the principal focus of the LAA standardization.

---

[1]Throughout the paper, we use the term Wi-Fi interchangeably with WLAN.

TABLE I
UNLICENSED BAND REGULATIONS BY REGION

| Region | 5.15-5.25 GHz | 5.25-5.35 GHz | 5.47-5.725 GHz | 5.725-5.85 GHz |
|---|---|---|---|---|
| USA | . | DFS/TPC | DFS/TPC | . |
| USA | FCC Part 15 Rules (Max EIRP, Emission Mask, etc.) | | | |
| EU [a] | Indoor only | Indoor only | Indoor/Outdoor | Indoor/Outdoor |
| EU [a] | . | DFS/TPC | DFS/TPC | . |
| EU [a] | ETSI Harmonized European Standards (LBT, Max EIRP, Emission Mask, etc.) | | | |
| China | Indoor only | Indoor only | TBD | Indoor/Outdoor |
| China | . | DFS/TPC | TBD | . |
| China | Max EIRP, Emission Mask, etc. | Max EIRP, Emission Mask, etc. | TBD | Max EIRP, Emission Mask, etc. |
| Japan | Indoor only | Indoor only | . | Band Not Available |
| Japan | . | DFS/TPC | DFS/TPC | Band Not Available |
| Japan | LBT, Max Burst Length (4ms), Max EIRP, Emission Mask, etc. | | | Band Not Available |
| Korea [b] | . | DFS/TPC | DFS/TPC | . |
| Korea [b] | Max EIRP, Emission Mask, etc. | | | |

N/A: Not applicable
LBT: Listen-Before-Talk
DFS: Dynamic Frequency Selection
TPC: Transmit Power Control
EIRP: Equivalent Isotropically Radiated Power
[a] EU band 4 is 5.725-5.875 GHz, where wireless access systems (WAS) are not operating in.
[b] Korea band 3 is 5.47-5.65 GHz.

The purpose of this article is to provide an overview of the LAA technology developed during the LTE Release 13. This article is organized as follows. In Section II, a brief overview is given on the relevant unlicensed spectrum bands and their associated regulatory requirements. This is followed by discussion on the deployment scenarios and use cases in Section III. Section IV summarizes the LAA standardization activities in 3GPP. We then take a deep-dive into the key technical features of the LAA in Section V. After that, we present a summary of evaluation results on verifying the coexistence between LAA and Wi-Fi, presented by a number of companies at 3GPP meetings. Finally, we draw conclusions in Section VII.

## II. UNLICENSED SPECTRUM AND REGULATIONS

The initial LAA deployments are expected to be limited to globally available 5 GHz unlicensed spectrum. Although the 5 GHz spectrum is generally designated as unlicensed spectrum, a radio equipment operating in the spectrum must abide by the regulatory requirements, which vary by regions as summarized in Table I. In addition to various requirements such as indoor-only use, maximum in-band output power, in-band power spectral density, and out-of-band and spurious emissions, the LTE operation in some unlicensed spectrum should also implement dynamic frequency selection (DFS) and transmit power control (TPC) depending on the operating band to avoid interfering with radars.

## III. SCENARIOS AND USE CASES

*A. Scenarios and Use Cases*

The introduction of carrier aggregation in LTE-Advanced required the distinction between a primary cell (PCell) and a secondary cell (SCell). The PCell is the main cell with which a user equipment (UE) communicates and maintains its connection with the network. One or more SCells can be allocated and activated to the UEs supporting carrier aggregation for bandwidth extension. Since the unlicensed carrier is shared by multiple systems, it can never match the licensed carrier in terms of mobility, reliability, and quality of service. Hence in LAA, the unlicensed carrier is considered only as a supplemental downlink (DL) SCell assisted by a licensed PCell via carrier aggregation. LAA deployment scenarios encompass scenarios with and without macro coverage, both outdoor and indoor small cell deployments, and both co-located and non-co-located (with ideal backhaul) cells operating in licensed and unlicensed carriers. Fig. 1 captured from 3GPP Technical Report (TR) 36.889 shows four considered LAA deployment scenarios [1].

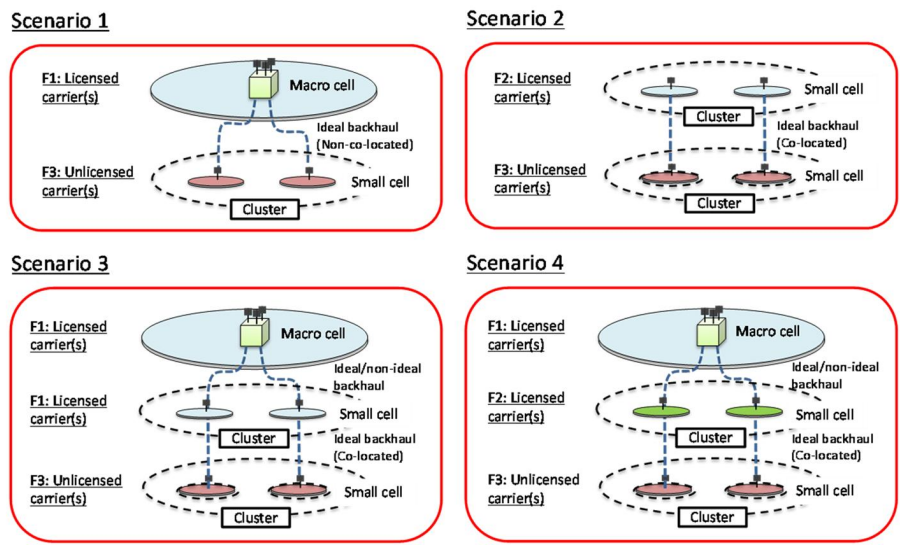

Fig. 1.  LAA deployment scenarios from 3GPP TR 36.889 [1]

- Scenario 1: Carrier aggregation between licensed macro cell (F1) and unlicensed small cell (F3).
- Scenario 2: Carrier aggregation between licensed small cell (F2) and unlicensed small cell (F3) without macro cell coverage.
- Scenario 3: Licensed macro cell and small cell (F1), with carrier aggregation between licensed small cell (F1) and unlicensed small cell (F3).
- Scenario 4: Carrier aggregation between licensed small cell (F2) and unlicensed small cell (F3). An ideal backhaul between macro cell and small cell can enable carrier aggregation between macro cell (F1), licensed small cell (F2) and unlicensed small cell (F3).

The carrier aggregation between non-co-located cells was mainly motivated by the hotspot scenario where a macro cell behaves as an anchor cell to provide robust connection management, while each small cell behaves as a booster cell to offer higher throughput. Such a hotspot scenario is the main use case of the small cells.

*B. Comparison of LAA with other LTE-based unlicensed technologies*

There are two competing LTE-based unlicensed technologies to LAA: LTE unlicensed (LTE-U) and LTE-WLAN Aggregation (LWA). LTE-U is based on the 3GPP Release 12 LTE technology to be used in the unlicensed spectrum. LTE-U uses adaptive on/off duty cycle as a mechanism to share the medium with existing Wi-Fi network. On the other hand, LWA is developed as part of Release 13 WI, which enables the simultaneous LTE and Wi-Fi connectivity via dual connectivity where the data traffic is aggregated at the eNB and routed to an operator core network.

When compared to the current implementation of Wi-Fi offloading, which prefers Wi-Fi connection regardless of the Wi-Fi link condition, the LTE-based unlicensed technologies anchored in the licensed carrier can provide better user experience thanks to reliable connection management and optimized link selection/activation. On the other hand, compared to LWA, LAA and LTE-U can provide a tighter integration of the licensed and unlicensed spectrum by performing lower layer aggregation, thereby improving the overall efficiency and especially the quality of delay-sensitive applications. Note however that not much technical details on LTE-U have been unveiled by the LTE-U Forum, which a closed organization consisting of a limited number of participating companies. On the other hand, LAA was standardized in 3GPP, whose resolution process is based on the consensus of all the participating companies, which also include companies representing Wi-Fi industry. As it will be seen in the following sections, the channel access mechanism of LAA largely resembles that of Wi-Fi and, thereby, it is natural to expect that LAA will provide better coexistence with existing Wi-Fi networks.

## IV. LAA IN RELEASE 13 AND BEYOND

The standardization of LAA in Release 13 was conducted in two phases; the first phase was the SI phase [2] and the second phase was the WI phase [3]. The goal of the SI phase was to study the feasibility of LTE enhancement to enable LAA operation in unlicensed spectrum while coexisting with other incumbent systems and fulfilling the regulatory requirements. The SI concluded that it is feasible for LAA to fairly coexist with Wi-Fi and other LAA networks, if an appropriate channel access scheme is adopted such as the listen-before-talk (LBT) [1], which is explained in details in Section V.A.

The main objective of the LAA WI is to specify LTE enhancements for operation in unlicensed spectrum, which is limited to support for LAA SCells operating with only DL transmissions, under the design criteria of a single global solution framework, fair coexistence between Wi-Fi and LAA, and fair coexistence between different LAA networks. The detailed objectives of the WI are to specify the support for the following functionalities: channel access framework including clear channel assessment, discontinuous transmission with limited maximum transmission duration, UE support for carrier selection, UE support for radio resource management (RRM) measurements including cell identification, time and frequency synchronization, channel-state information (CSI) measurement. The LAA WI was completed by the end of 3GPP Release 13 in late 2015.

## V. KEY TECHNICAL FEATURES OF THE RELEASE 13 LAA

### A. LBT and Overall DL Data Transmission

LBT is a procedure whereby radio transmitters first sense the medium and transmit only if the medium is sensed to be idle, which is also called clear channel assessment (CCA). The CCA utilizes at least energy detection (ED) to determine the presence of signals on a channel. Recall from Section II that LBT in 5 GHz unlicensed spectrum is required in Europe and Japan but not in US, China, and Korea. However, the adoption of LBT is necessary for LAA to become a single global solution that complies with any regional regulatory requirements. Apart from regulatory requirements, LBT is highly beneficial for fair and friendly coexistence with incumbent systems in the unlicensed spectrum and with other LAA networks. The main incumbent systems in the 5 GHz band are the WLANs based on IEEE 802.11n/ac technologies, which are widely deployed both by individuals and operators for data offloading. The WLAN employs contention-based channel access mechanism, called carrier sense multiple access with collision avoidance (CSMA/CA) [4]. A WLAN node intends to transmit first performs CCA before transmission. Additional backoff mechanism is designed for the collision avoidance aspect to cope with the situation when more than one node senses the channel idle and transmits at the same time. The backoff counter is drawn randomly within the contention window size (CWS), which is increased exponentially upon the occurrence of collision and reset to the minimum value when the transmission succeeds.

The LBT mechanism designed for LAA fundamentally resembles the CSMA/CA of WLAN. The specified LBT procedure for LAA DL transmission bursts containing PDSCH[2], is illustrated in Fig. 2. The size of the LAA contention window is variable between X and Y extended CCA (ECCA) slots, which are the minimum and maximum CWSs. The ECCA slot duration is at least 9 μs, which is exactly the same as WLAN slot.

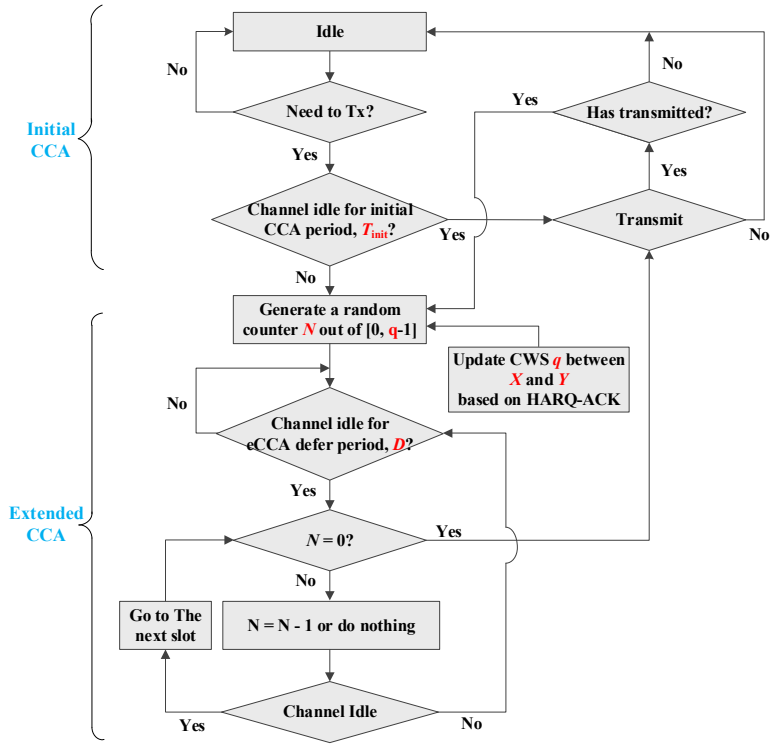

Fig. 2. Flowchart of the recommended DL LBT procedure by 3GPP

Illustration of an LAA DL burst transmission is given in Fig. 3, where MCOT stands for maximum channel occupancy time. 3GPP has introduced four different priority classes for DL LBT with random backoff, where the smaller the LBT priority class number, the higher the priority. Release 13 supports at least priority class 3 and best effort traffic shall not use a priority class with higher priority than the priority class 3. 3GPP has differentiated the MCOT according to the LBT priority classes. For priority classes 3 and 4, MCOT is 10 ms, if the absence of any other technology sharing the carrier can be guaranteed on a long term basis. Otherwise, it is 8 ms. For LAA operation in Japan, the E-UTRAN NodeB (eNB) may need to sense the channel to be idle for additional single continuous interval of duration 34 μs after every 4ms of transmission if the DL transmission burst is longer than 4 ms.

*1) ED Threshold*

An important component of LBT design is the choice of ED threshold, which determines the level of sensitivity to declare the existence of ongoing transmissions. 3GPP considers the mechanism to adapt the ED threshold. For instance, if the absence of any other technology sharing the carrier cannot be guaranteed on a long term basis (e.g., by level of regulation), the maximum energy detection threshold used by LAA for category 4 LBT is

$$TH = \max(-72 \text{ dBm (20MHz)}, \min(T_{max}, T_{max} - 10 \text{ dB} + (P_H - P_{TX})))),$$

where $P_H$ is a reference power equaling 23 dBm, $P_{TX}$ is the configured maximum transmit power for the carrier in dBm, and it is given by $T_{max}$ = -75 dBm/MHz+ 10*log10($BW_{MHz}$) where $BW_{MHz}$ is the channel bandwidth in MHz. In a nutshell, the ED threshold can be raised if the bandwidth $BW_{MHz}$ becomes wider and/or the transmit power $P_{TX}$ is lowered.

---

[2] PDSCH is the Physical Downlink Shared Channel in LTE for the transmission of unicast user data and paging information.

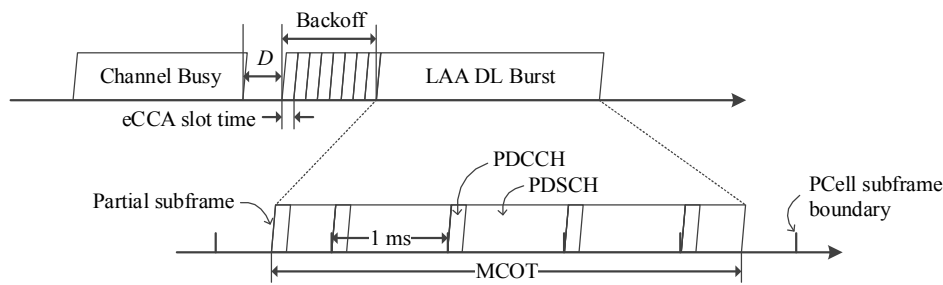

Fig. 3. Illustration of an LAA DL burst transmission

*2) CWS Adaptation*

The CWS is adapted based on the hybrid automatic repeat request (HARQ)-ACK feedback[3]. HARQ-ACK feedback can take a value from ACK, NACK, and DTX, where ACK refers to the situation of correct reception, NACK refers to the situation where control information (i.e., PDCCH[4]) is correctly detected but there is an error in the data (i.e., PDSCH) reception, and DTX refers to the situation when a UE misses control message containing scheduling information (i.e., PDCCH), rather than the data itself (i.e., PDSCH). The set of CWSs for LAA DL transmission bursts containing PDSCH for priority class 3 is {15, 31, 63}, i.e., X and Y in Fig. 2 are 15 and 63, respectively. The X and Y values are set differently for different LBT priority classes. The CWS is increased if at least 80% of the HARQ-ACK feedback values for the first subframe of a DL burst are NACK. The CWS increase is in an exponential manner as in Wi-Fi. Otherwise, the CWS is reset to the minimum value. DTX is considered as NACK except when the UEs were not actually scheduled by eNB or the scheduling information was sent through the licensed PCell.

*3) Multicarrier LBT*

LAA supports two alternative solutions for multi-carrier LBT. In the first option, the eNB is required to designate a carrier requiring LBT with random backoff as illustrated in Fig. 2 and the eNB can sense other configured carriers with single interval LBT only if the eNB completes the LBT with random backoff on the designated carrier. In the second option, the eNB performs LBT with random backoff on more than one unlicensed carriers and is allowed to transmit on the carriers that has completed the LBT with potential self-deferral to align transmissions over multiple carriers.

*B. LBT for LAA Discovery Reference Signal*

In LTE Release 12, discovery reference signal (DRS) was introduced to facilitate fast transition of small cell from OFF state to ON state by transmitting low duty cycle signals for radio resource management (RRM) measurement during OFF state. During the OFF period, DRS, consisting of synchronization signals and reference signals, is transmitted to allow UEs to discover and measure the dormant cell. The RRM measurement details are further explained in Section V.C.2. LAA DRS is the same as the first twelve OFDM symbols of the Release 12 DRS in Frame Structure Type 1[5] (frame structure defined for frequency division duplexing (FDD)).

DRS can be transmitted within a periodically occurring time window called DRS measurement timing configuration (DMTC) occasion which has a duration of 6 ms and a configurable period of 40/80/160 ms. The transmission of DRS is also subject to LBT.

---

[3] HARQ is a mechanism works at physical layer to deal with the errors in the reception of transmitted data. Unlike the ARQ, the retransmission in response to the occurrence of errors are different redundancy versions of the original coded block.
[4] PDCCH refers to the Physical Downlink Control Channel used to convey DL control information, including DL/UL scheduling grants.
[5] The 1 ms LTE subframe consists of two slots of 0.5 ms each. Each slot contains either six or seven OFDM symbols, depending on the cyclic prefix (CP) length. LAA supports only normal CP, which corresponds to seven OFDM symbols per slot.

A DL transmission burst containing DRS without PDSCH follows a single idle observation interval of at least 25 µs. Due to LBT, the DRS may not be transmitted as frequent as scheduled. To increase the DRS transmission opportunity so as to improve the performance of functionalities (e.g., synchronization, RRM measurement) relying on DRS, DRS can be transmitted by the network once in any subframe within the DMTC occasion.

*C. LTE Enhancement to Support LAA*

*1) Frame Structure and Partial Subframe*

For LAA, the LBT procedure can be completed at any time. Moreover, DL transmission may not start/end at the subframe boundary. To support such flexible operation for LAA, a new frame structure, called Type 3, has been introduced in Release 13 for which UE considers each subframe as empty unless DL transmission is detected in that subframe.

Note that other neighboring system can take the transmission opportunity while the LAA eNB is awaiting the next subframe boundary unless a reservation signal is transmitted after successful LBT. To efficiently utilize the radio resources, partial subframe has been introduced for LAA SCell, where DL transmission, excluding reservation signal, can start at the first or second slot boundaries of a subframe as illustrated in Fig. 3. Depending on starting position of DL transmission and due to MCOT limitation, DL transmission may not end at the subframe boundary. To utilize the ending partial subframe with minimal specification efforts, the existing Downlink Pilot Time Slot (DwPTS) structure is reused, where DwPTS is the DL portion of the special subframe of the Frame Structure Type 2 for time division duplexing (TDD). With the existing DwPTS configuration, the duration of the last subframe of a DL transmission burst can be one of {3, 6, 9, 10, 11, 12} OFDM symbols or a whole subframe consisting of 14 OFDM symbols. Common control signal in LAA SCell is used to indicate the number of OFDM symbols of the current and the next subframe for DL transmission.

*2) RRM Measurement*

RRM measurement is required for proper LAA SCell selection/reselection. RRM measurement is based on the reception of DRS containing CRS/CSI-RS[6] and the reporting consists of Reference Signal Received Power (RSRP) and Reference Signal Received Quality (RSRQ)[7]. Due to the dynamically changing channel condition in unlicensed spectrum, legacy RRM measurement reporting may not be sufficient to reflect load conditions, interference outside DL burst, and potential hidden nodes in the unlicensed channel. In this regard, LAA UEs can be configured to report average RSSI and channel occupancy as a part of RRM measurements. Average RSSI provides an estimation of load conditions and captures the overall interference on LAA SCell. The channel occupancy is defined as the percentage of time when the channel is sensed to be busy, i.e., when the measured RSSI sample is above a predefined threshold. It is important that the Layer 1 (L1) averaging duration of UE-reported RSSI measurement should roughly be of the same order as the minimum transmission granularity on an unlicensed carrier. For example, Wi-Fi ACK duration can typically be less than 100 µs. As a result, the L1 averaging duration is one LTE OFDM symbol. In addition, multiple consecutive L1 RSSI samples can be aggregated to produce measurement durations ranging from 1 ms to 5 ms.

*3) Cell Detection and Synchronization*

Cell detection and synchronization rely on the reception of the synchronization signals such as primary and secondary synchronization signal (PSS/SSS) and CRS. Specifically, PSS/SSS can be used for physical-layer cell identity (ID) detection, and CRS can be used to further improve the performance of cell ID detection, for example, to confirm cell detection. PSS/SSS and CRS can also be used to acquire coarse and fine time/frequency synchronization, respectively. Thanks to the multiple DRS transmission opportunities within a DMTC occasion, large time/frequency drift between two successive DL bursts is unlikely. Thus, the synchronization based on DRS in LAA systems can achieve reliable performance. On the other hand, the DL subframe presence detection by UE is needed as the eNB does not always transmit. The exact detection method employed is left to UE implementation.

---

[6] CRS and CSI-RS refer to cell-specific reference signal and channel state information reference signal, respectively.
[7] RSRQ is calculated as the ratio of RSRP to Reference Signal Strength Indicator (RSSI), indicating the ratio of the received signal power to the total received power including its own signal power, interference and noise.

*4) CSI Measurement and Reporting*

LAA supports transmission modes (TMs) with CRS-based CSI feedback, including TM1, TM2, TM3, TM4 and TM8, and those with CSI-RS based CSI feedback, including TM9 and TM10[8]. CSI-RS/CSI-IM[9] for CSI measurement is present in the configured periodic CSI-RS/CSI-IM subframes within DL transmission bursts. Similar to the legacy LTE systems, both periodic and aperiodic CSI reports are supported. Unlike the legacy LTE system whereby the CRS/CSI-RS transmission power, or Energy Per Resource Element (EPRE), is fixed, CRS/CSI-RS transmission power on LAA SCell is only fixed within a DL transmission burst while it can vary across DL transmission bursts. As a result, UE should not average CRS/CSI-RS measurements across transmission bursts. UE could either rely on CRS detection or common control signaling to differentiate DL bursts.

*5) Scheduling and HARQ*

LTE supports two different scheduling approaches, namely cross-carrier scheduling and self-scheduling. With cross-carrier scheduling, the control information including scheduling indication, i.e., PDCCH, and the actual data transmission, i.e., PDSCH, take place on different carriers, whereas they are transmitted on the same carrier in the case of self-scheduling. Due to the uncertainty of channel access opportunities on unlicensed carriers, the synchronous HARQ protocol[10] with fixed time relation between retransmissions, is difficult to use for LAA. Thus, the existing asynchronous HARQ protocol can be used for LAA DL/UL. For LAA UL, in particular, UEs would need to rely on the UL grant from eNB for UL (re)transmissions.

## VI. COEXISTENCE PERFORMANCE

*A. Evaluation Methodology*

The first and foremost goal of LAA design is to ensure the fair coexistence with other incumbent systems operating in the same unlicensed spectrum. This is captured in the LAA design target in terms of fair sharing metrics; an LAA network should not impact Wi-Fi services more than an additional Wi-Fi network on the same carrier. In this section, we highlight the extensive evaluation efforts contributed by numerous sources during the LAA SI phase [1].

3GPP defined indoor scenario consists of four equally spaced LAA eNBs and/or Wi-Fi APs deployed by each operator in a single story building serving 10 uniformly distributed LAA UEs and/or Wi-Fi STAs operating on the same unlicensed carrier. The distance between two closest nodes from two operators is random. The set of small cells for both operators is centered along the longer dimension of the building. Outdoor scenario considers a hexagonal grid with 3 sectors per site and inter-site distance of 500m. Clusters of small cells are uniformly random within macro geographical area. Within each cluster, there are 4 small cells per operator, randomly dropped within cluster area. 10 UEs are randomly dropped within coverage area of small cell in unlicensed spectrum. 3GPP also considered various traffic models such as File Transfer Protocol (FTP) traffic, and mixed FTP and Voice over Internet Protocol (VoIP) traffic. The Wi-Fi network with DL only traffic and both DL and UL traffic were considered as well. To verify the coexistence, a two-step methodology is used; in step 1, the performance of two coexisting Wi-Fi networks is evaluated as a benchmark and, then, in step 2, a Wi-Fi network is replaced with an LAA network and the performance of the non-replaced

---

[8] The TMs differ in terms of number of antennas, MIMO mode, and number of spatial streams, etc.
[9] CSI-IM refers to channel-state information – interference measurement, whose resource configuration is based on zero-power CSI-RS configuration.
[10] HARQ protocol can be categorized into synchronous and asynchronous HARQ based on the flexibility in the time domain. With synchronous HARQ, re-transmission occurs at fixed time, while with asynchronous HARQ, re-transmission can occur at any time.

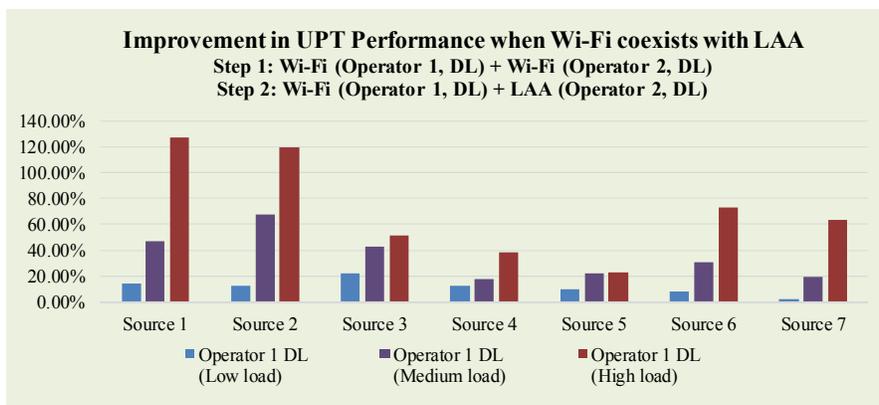

Fig. 4. Improvement in the UPT for the DL only Wi-Fi network (Sources 1-7 are from 3GPP contributions R1-150694, R1-152732, R1-151821, R1-152863, R1-153384, R1-153426, and R1-153629, respectively.)

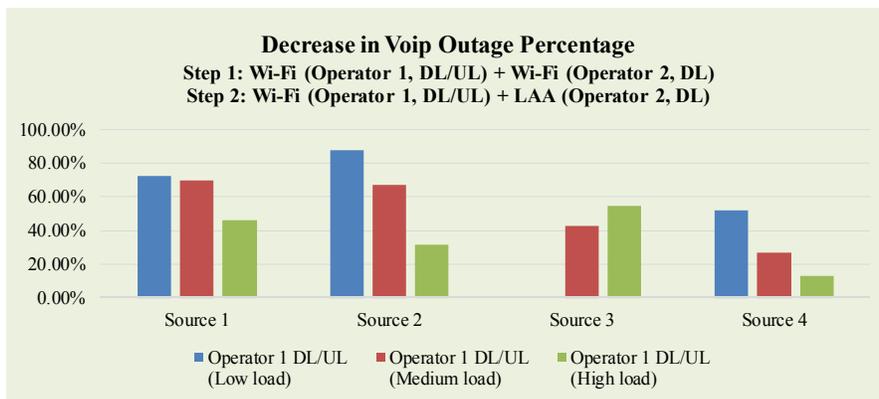

Fig.5. Decrease in VoIP outage for the DL/UL Wi-Fi network (Sources 1-4 are from 3GPP contributions R1-152326, R1-152642, R1-152937, and R1-153343, respectively.)

Wi-Fi network is compared against step 1.

*B. Evaluation Results from 3GPP*

During the discussion in 3GPP, it was identified that ensuring coexistence for indoor scenario is more difficult than that for outdoor scenario due to close proximity between LAA eNBs and Wi-Fi access points (APs)/stations (STAs). It is also apparent that restricting LAA eNB to transmit data only in the unlicensed carrier is more challenging to prove fair coexistence because the licensed carrier given to LAA eNB is an additional resource that can be exploited to alleviate the transmission demand on unlicensed spectrum in step 2, resulting in a more friendly environment for fair coexistence. The results captured in this section from [1] are thus focused on the most demanding scenarios in proving the fair coexistence. The IEEE 802.11ac technology is assumed for Wi-Fi networks.

The user perceived throughput (UPT) is considered by 3GPP as an important performance measure for network serving non-full-buffer traffic. The UPT is defined as the amount of data over the actual time spent for downloading excluding the idle time for waiting files to arrive. Fig. 4 shows the improvement in the UPT for the non-replaced DL only Wi-Fi network in step 2 compared to step 1 with different loading conditions. Buffer occupant time of 15-30%, 35-50%, and 60-80% averaged over APs of the non-replaced Wi-Fi network in step 1 is considered as low, medium, and high load, respectively. From Fig. 4, it can be observed that the Wi-Fi UPT performance is improved when the Wi-Fi network coexists with an LAA network rather than another Wi-Fi network. This is mainly because LTE has higher spectral efficiency than Wi-Fi due to the better link adaptation based on explicit CSI feedback, while the control messages such as CSI feedback can go through licensed carrier. Consequently, the interference from Operator 2 to Operator 1 is reduced in step 2, thereby, improving the Wi-Fi performance in step 2. Fig. 5 shows the coexistence performance when Operator 1's Wi-Fi network serves bidirectional, i.e., both DL and UL, mixed FTP and VoIP traffic. From the figure, it is shown that VoIP outage for non-replaced Wi-Fi network can be reduced significantly when it coexists with LAA network. This draws the conclusion that 3GPP LAA design can indeed ensure the coexistence with incumbent Wi-Fi networks for both non-real-time and real-time traffic.

Finally, we make a note on the observations made during the early SI phase. Multiple sources identified that the Wi-Fi performance can be significantly degraded when it coexists with LAA. These observations were the main motivation behind the adoption of the LBT algorithm based on the exponential backoff as in Wi-Fi. The simulation results summarized here are those following the LBT algorithm, which was finally agreed.

## VII. CONCLUSION

This article gave an overview of 3GPP Release 13 LAA technology. The LAA supplements a licensed primary carrier with unlicensed secondary carriers via carrier aggregation. The 3GPP aimed at not only meeting the regulatory requirements but also ensuring fair coexistence with existing Wi-Fi networks. These design goals have led to significant changes at the LTE physical layer for LAA. Based on the evaluations contributed to 3GPP provided from a wide spectrum of sources, there is a consensus that LAA can fairly coexist with Wi-Fi networks serving various traffic types.

## REFERENCES

[1] Study on Licensed-Assisted Access to Unlicensed Spectrum, 3GPP TR 36.889, May. 2015.
[2] RP-141664 "Study on Licensed-Assisted Access using LTE," Ericsson, Qualcomm, Huawei, Alcatel-Lucent, 3GPP TSG RAN Meeting #65, Sep. 2014.
[3] RP-151045 "New Work Item on Licensed-Assisted Access to Unlicensed Spectrum," Ericsson, Huawei, Qualcomm, Alcatel-Lucent, 3GPP TSG RAN Meeting #68, Jun. 2015.
[4] Part 11: Wireless LAN Medium Access Control (MAC) and Physical Layer (PHY) Specifications, IEE Std 802.11™-2012, Mar. 2012.